\documentclass[conference]{IEEEtran}
\IEEEoverridecommandlockouts

\usepackage{cite}
\usepackage{amsmath,amssymb,amsfonts}
\usepackage{algorithmic}
\usepackage{graphicx}
\usepackage{textcomp}
\usepackage{xcolor}

\newcommand{\x}{\mathbf{x}}
\newcommand{\ct}{\mathbf{c}}
\newcommand{\z}{\mathbf{z}}
\newcommand{\q}{\mathbf{q}}

\usepackage{multirow}
\usepackage{mathrsfs}
\usepackage[utf8]{inputenc}

\def\BibTeX{{\rm B\kern-.05em{\sc i\kern-.025em b}\kern-.08em
    T\kern-.1667em\lower.7ex\hbox{E}\kern-.125emX}}
\begin{document}

\title{Fairness and Privacy in Voice Biometrics:\\ A Study of Gender Influences Using wav2vec 2.0
}

\makeatletter
\newcommand{\linebreakand}{%
  \end{@IEEEauthorhalign}
  \hfill\mbox{}\par
  \mbox{}\hfill\begin{@IEEEauthorhalign}
}
\makeatother

\author{
  \IEEEauthorblockN{Oubaïda Chouchane, Michele Panariello, Chiara Galdi, Massimiliano Todisco, Nicholas Evans}\\
    EURECOM\\
    Sophia Antipolis, France \\ \\
    \textit{firstname} [dot] \textit{lastname} [at] \textit{eurecom} [dot] \textit{fr}}

\maketitle

\begin{abstract}
This study investigates the impact of gender information on utility, privacy, and fairness in voice biometric systems, guided by the General Data Protection Regulation (GDPR) mandates, which underscore the need for minimizing the processing and storage of private and sensitive data, and ensuring fairness in automated decision-making systems. We adopt an approach that involves the fine-tuning of the wav2vec 2.0 model for speaker verification tasks, evaluating potential gender-related privacy vulnerabilities in the process. Gender influences during the fine-tuning process were employed to enhance fairness and privacy in order to emphasise or obscure gender information within the speakers' embeddings.
Results from VoxCeleb datasets indicate our adversarial model increases privacy against uninformed attacks, yet slightly diminishes speaker verification performance compared to the non-adversarial model. However, the model's efficacy reduces against informed attacks. Analysis of system performance was conducted to identify potential gender biases, thus highlighting the need for further research to understand and improve the delicate interplay between utility, privacy, and equity in voice biometric systems.
\end{abstract}


\section{Introduction}
The voice is an appealing approach to biometric authentication. Its merits include ease of use, contactless and natural interaction, efficiency, and application to authentication at a distance, e.g.\ over the telephone. However, the voice is a rich source of personal information and recordings of speech can be used to infer far more than just the speaker's identity, e.g.\ the speaker's gender\cite{Za21}, ethnicity\cite{HRC13}, and health status\cite{SLRR21}. The safeguarding of such extraneous personal information is nowadays essential; without it, there is no guarantee that recordings of speech will not be used for purposes beyond person authentication\cite{SDAA19}.

The General Data Protection Regulation (GDPR)\footnote{https://gdpr-info.eu/} calls for adequate protections for personal data, encompassing both \textit{sensitive }biometric information like voice and \textit{personal} attributes such as gender\footnote{https://www.gdpreu.org/the-regulation/key-concepts/personal-data/}. In adherence to Art. 4(1) of the GDPR, personal data processing must abide by principles of legality and fairness, managing data in line with reasonable expectations and avoiding unjust harm. Any AI-driven data processing resulting in unfair discrimination violates this principle.

As mandated by GDPR, this study particularly emphasizes privacy and fairness, focusing on gender due to its demonstrated influence on speaker authentication services\cite{HD22} and the observed gender bias in voice assistant responses\cite{Li19}. GDPR aims to protect the rights and freedoms of individuals, including privacy and non-discrimination, with regard to personal data processing. 
Concealing gender adheres to the principles of data minimization and privacy by design, limiting the risk of misuse or unauthorized data access.

In this research, we grapple with the triple challenge of utility, privacy, and fairness in speaker verification systems. Starting with fine-tuning a pre-trained wav2vec 2.0 for speaker verification tasks, we then evaluate potential vulnerabilities tied to gender privacy and the fairness of Automatic Speaker Verification (ASV) performance across genders. Subsequently, we implement an adversarial technique during the fine-tuning process to conceal gender information in the speaker embeddings, thereby enhancing user privacy.
To conclude, we present a comprehensive analysis of the impact of gender information on the utility, privacy, and fairness of the systems we propose.

\section{Related work}
Significant strides have been made in speaker verification, with efforts concentrated on enhancing user privacy. These strategies prioritize the protection of gender-specific data without sacrificing system utility. Noé et al.\cite{No20} suggested an Adversarial Auto-Encoder (AAE) method to separate gender aspects from speaker embeddings while preserving ASV performance. The approach uses an external gender classifier to analyze encoded data. Later, they leveraged a normalizing flow to control gender information in a flexible manner\cite{No22}. In another study, Benaroya et al.\cite{BOR21} developed a novel neural voice conversion framework using multiple AEs to create separate linguistic and extra-linguistic speech representations, allowing adjustments during the voice conversion process. Recently, Chouchane et al.\cite{Ch23} used an adversarial approach to hide gender details in speaker embeddings while ensuring their effectiveness for speaker verification. They incorporated a Laplace mechanism layer, introducing noise to obscure gender information and offering differential privacy during inference. 

In terms of fairness, research reveals a distinct disparity in ASV system performance based on gender, exposing gender bias\cite{TD21}. Two primary strategies to mitigate this bias include pre-processing and in-processing. Pre-processing uses balanced datasets for training, as Fenu et al.\cite{Fe20} demonstrated with gender, language, and age-balanced data. In contrast, in-processing infuses fairness directly during training, as seen in Shen et al.'s Group-Adapted Fusion Network (GFN)\cite{Sh22} and Jin et al.'s adversarial re-weighting (ARW) approach\cite{Ji22}. Peri et al.\cite{PSN23} recently proposed adversarial and multi-task learning techniques for bias mitigation, highlighting a potential trade-off between system utility and fairness. 

Finally, shifting focus to system utility, a cornerstone in ASV performance, the wav2vec 2.0\cite{Ba20}, a self-supervised framework for speech representation learning, enters the scene. The wav2vec 2.0 can be effectively adapted for speaker verification tasks \cite{VVL22,Fa20}.

\section{Automatic speaker verification, gender recognition and suppression using wav2vec 2.0}
In this section, we outline our use of the wav2vec 2.0 model, a versatile speech feature encoder that is pre-trained through self-supervision and can be adapted to specific tasks.
We fine-tuned wav2vec 2.0 for three distinct tasks: speaker recognition, and gender recognition and suppression. Section 3.1 elaborates on the pre-training process, while Section 3.2 details our contributions to fine-tuning.
Both procedures are graphically depicted in Fig.~\ref{fig:system}.

\subsection{Pre-training}
Given a raw audio input signal $x$, wav2vec 2.0 produces a set of $T$ feature vectors $\ct_1,\dots,\ct_T$.
The model is split into a 1D-convolutional encoder and a Transformer module\cite{Va17} two main parts.
First, the encoder maps the raw audio $\x$ to latent feature vectors $\z_1,\dots,\z_T$. 
The latent features are then fed into the Transformer module to produce output feature vectors $\ct_1,\dots,\ct_T$, and are also used to compute a set of quantised macro-codewords $\q_1,\dots,\q_T$.
Each macro-codeword $\q_t$ is the concatenation of $G$ codewords $\q_{t,1},\dots,\q_{t,G}$ selected from $G$ different codebooks $\mathcal{Q}_1,\dots,\mathcal{Q}_G$, each of size $V$, learned at training time.
Each codeword $\q_{t,j}$ is sampled from $\mathcal{Q}_j$ according to a $V$-fold categorical distribution. The distribution is optimized during pre-training and computed as $\mathbf{p}_{t,j} = \text{GS}(\z_t)$, where GS indicates a linear layer projecting $\z_t$ to $V$ dimensions followed by a straight-through Gumbel-softmax estimator\cite{JGP17}.

During pre-training, the model attempts to simultaneously minimize a \emph{contrastive} loss $\mathcal{L}_m$ and a \emph{diversity} loss $\mathcal{L}_d$. To compute the former, some of the latent feature vectors $\z_1,\dots,\z_T$ are randomly masked. 
Then, for each masked $\z_t$, the Transformer module attempts to compute $\ct_t$ so that it is as similar as possible to the corresponding quantised macro-codeword $\q_t$, and as dissimilar as possible from other ``distractor'' macro-codewords $\Tilde{\q}$ randomly sampled from the rest of the batch. The quantised macro-codewords are computed with no masking.
The \emph{diversity} loss $\mathcal{L}_d$ encourages the model to make uniform use of all the $V$ codewords in each codebook by maximizing the entropy of the average probability distribution $\Bar{\mathbf{p}}_g$ produced by all $\z_t$ in a batch for each codebook $g$.
The overall loss is:
\begin{equation}
    \mathcal{L} = 
    \;\;\;
    \underbrace{
        - \sum_{\substack{\text{masked}\\\text{steps }t}} \log \frac
        {
            \exp\left(s(\ct_t, \q_t) / \kappa \right)
        }
        {
            \sum_{\Tilde{\q}} \exp\left(s(\ct_t, \Tilde{\q}) / \kappa \right)
        }
    }_{\mathcal{L}_m}
    \;\;\;
    \underbrace{
        - \alpha \frac{1}{GV} \sum_{g=1}^{G} H\left(\Bar{\mathbf{p}}_g\right)
    }_{\mathcal{L}_d}
\end{equation}
Where $\kappa$ is a temperature coefficient, $s$ is the cosine similarity, $\alpha$ is a weight hyperparameter and $H$ indicates entropy.

\subsection{Fine-tuning for speaker verification and gender recognition}
\label{sec:fine-tuning}
In this paper, we fine-tune wav2vec 2.0 for the downstream tasks of speaker verification and gender recognition.
In both cases, for each input utterance $\x$, the output features $\ct_1, \dots, \ct_T$ are averaged across time to obtain a 1-dimensional embedding $\ct$.
In the case of gender recognition, $\ct$ is then passed through a linear layer $f_g$ which is trained by optimising the cross-entropy loss $\mathcal{L}_g$ between the predicted logits and the true gender label for each utterance (0 for male, 1 for female).
For speaker verification, $\ct$ is passed through a different linear layer $f_{s}$ of $N$ output neurons, where $N$ is the number of speakers in the training dataset. The layer is then optimized to perform speaker identification by minimizing the additive angular margin (AAM) softmax loss $\mathcal{L}_{s}$\cite{Xi19}. At test time, the final embedding $\ct$ is used as a trial or enrollment vector.
Overall, the final loss can be formulated as:
\begin{equation}
    \label{eq:loss}
    \mathcal{L} = \lambda \mathcal{L}_{s} + (1 - \lambda) \mathcal{L}_{g}
\end{equation}
where $\lambda$ is a hyper-parameter between 0 and 1 that controls the weight of each loss component. We experimented with three different model configurations:
Model 1 ($M_s$) is fine-tuned for speaker verification, i.e.\ $\lambda = 1$;  Model 2 ($M_{sg}$) is fine-tuned for both tasks, i.e.\ $\lambda = 0.5$; Model 3 ($M_{sga}$) is optimised in a similar manner, though with a gradient reversal layer\cite{Ga16} $g_r$ to suppress gender information.

The optimization process becomes an adversarial game between $f_g$, which attempts to minimize $\mathcal{L}_g$, and the backbone, which attempts to maximize it. Meanwhile, the $\mathcal{L}_s$ component is optimized as usual.

\begin{figure*}[!t]
    \centering
    \includegraphics[width=\textwidth]{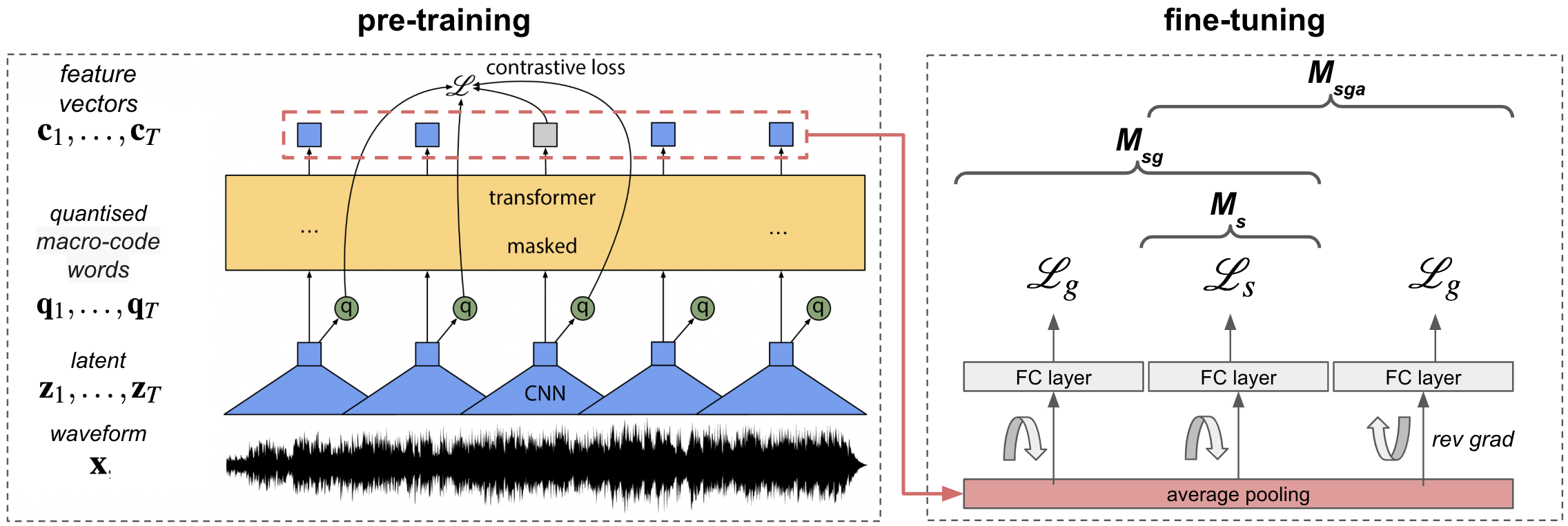}
    \caption{Graphical depiction of the proposed systems. $M_s$: fine-tuning the speaker identification task. Msg: fine-tuning gender and speaker identification. $M_{sga}$: similar to $M_{sg}$, but the gender identification task is made adversarial.}
    \label{fig:system}
\end{figure*}

\section{Experimental setup}
Described in this section are the databases used for all experimental work, the metrics used for evaluation, and the fine-tuning procedure.

\subsection{Databases}
We used the VoxCeleb1 and VoxCeleb2 speaker recognition databases\cite{NCZ17,CNZ18}.
VoxCeleb1 includes over 100,000 utterances from 1,251 celebrities, while VoxCeleb2 contains over a million utterances from 6,112 speakers. Both datasets, compiled from YouTube videos, are widely used for speaker recognition and voice-related machine-learning tasks.
Fine-tuning is performed using the VoxCeleb2 development set which contains data collected from 5994 unique speakers of which 3682 are male and 2312 are female, corresponding to an imbalance in favour of male speakers of 22.9\%. To assess the performance of our systems, we used the VoxCeleb1 test set, which consists of 40 unique speakers of which 25 are male and 15 are female.

\subsection{Metrics}
A range of key metrics was selected, many of which are derived from the evaluation of biometric classification systems, e.g. speaker verification and gender classification. The following describes how they are used to jointly assess the utility, privacy, and fairness of the models under scrutiny.\\

\textbf{Utility} is measured by assessing the performance for the task of automatic speaker verification (ASV) in terms of equal error rate (EER). EER is the operating point defined by the detection threshold $\tau$ at which the false acceptance rate~(FAR) and the false rejection rate~(FRR) are equal.\\

\textbf{Privacy} relates to the difficulty of an adversary to infer sensitive attributes. 
We use AUC (area under the receiver operating characteristic curve) metric to gauge privacy. In contrast to EER, AUC provides a comprehensive view, which is ideal for evaluating system security across diverse threshold selections.\\

\textbf{Fairness} is aimed at ensuring that a system behaves equally with all subgroups of the target population. Many approaches for measuring fairness have been proposed recently and there is still no agreement on which is the most appropriate. We adopted two different metrics with the aim of giving a more meaningful insight into the fairness of the models.\\

The first adopted approach aims at ensuring that the error rates for all demographic groups fall within a small margin $\epsilon$. However, for practical purposes, given a pair of demographic groups $D = {d_1, d_2}$, we calculate $A(\tau)$ and $B(\tau)$, as:

\begin{equation}
    A(\tau)= max\left ( \left | FAR^{d_1}(\tau) - FAR^{d_2}(\tau)\right | \right ) 
\end{equation}
\begin{equation}
    B(\tau)= max\left ( \left | FRR^{d_1}(\tau) - FRR^{d_2}(\tau)\right | \right ).
\end{equation}

These represent the maximum absolute differences in FAR and FRR across all groups.
In a perfect system, both $A(\tau)$  and $B(\tau)$  would equal 0, reflecting identical error rates across all groups. 

The Fairness Discrepancy Rate (FDR)\cite{dFPM21} is defined as:

\begin{equation}
    FDR(\tau)= 1 - (\alpha A(\tau) + (1 - \alpha)B(\tau))
    \label{eq:fdr}
\end{equation}

where the hyper-parameter $\alpha \in [0, 1]$ determines the relative importance of false alarms.
FDR ranges between 0 and 1 and would equal 1 in the case of a perfectly fair system.
However, achieving perfect fairness is often unrealistic, leading to the introduction of $\epsilon$ which allows for certain discrepancies. Though $\epsilon$ isn't included in the FDR calculation, it's vital for defining an acceptable level of fairness and interpreting FDR results.\\
Given the absence of a universal $\epsilon$ and the complexities of biometrics, absolute fairness often isn't achievable. Thus, FDR and Area Under FDR (auFDR) are used to compare the fairness of different biometric systems. The auFDR is calculated by integrating the FDR over a specific threshold range $\tau$, denoted as $FAR_x$. To fairly compare the auFDR between different systems, the specific range of $\tau$ used must be reported, as the value of the auFDR depends on this range. Like the FDR, the auFDR varies from 0 to 1, with higher values denoting better fairness. In our experiments, we set the range to FARs below 0.1; FARs above this value correspond to a system with little practical interest.\\

The second metric is the fairness activation discrepancy~(FAD), which we use to investigate fairness \textit{within} the network. FAD is inspired by \emph{InsideBias}\cite{Se21}, a fairness metric developed originally for the study of face biometrics and which we adapt to our study of voice biometrics. Notably, this adaptation of FAD for voice biometrics is a novel metric in this context.

\emph{InsideBias} is based upon the examination of neuron activations and the comparison of model responses to demographic groups within distinct layers. In\cite{Se21}, the authors observed that underrepresented groups corresponded to lower average activations.
In the case of voice biometrics, the output of each network layer can be viewed as a bi-dimensional tensor of neurons over temporal frames:

\begin{equation}
    A_{ij}^{[l]} = \Psi^{[l]}(\cdot)
\end{equation}

where $i=1,..., N$, $j=1,..., M$, \(A_{ij}\) is the activation of the $i^{th}$ neuron for the $j^{th}$ temporal frame, \(\Psi^{[l]}\) is the activation function at layer \(l\), and \(N\) and \(M\) are the total number of neurons and frames respectively.
For each layer $l$ we calculate the root mean square of $A_{ij}$ over the $j^{th}$ frame which serves to account for large positive or negative activations. Then, we take the maximum along the $i^{th}$ feature dimension:

\begin{equation}
    {\Lambda}^{[l]} = \max_i\sqrt{\left( \frac{1}{M} \sum_{j}^{} A_{ij}^2\right)}
\end{equation}

The FAD is defined as the absolute difference between $\Lambda$ for a pair of two distinct groups and is given by $FAD = |\Lambda_{d_1} - \Lambda_{d_2}|$. Near-zero values of FAD indicate better fairness.

\subsection{Fine-tuning procedure}
$M_s$, $M_{sg}$ and $M_{sga}$ models are fine-tuned as described in Section~\ref{sec:fine-tuning}.
An initial warm-up is applied to the linear classification heads for the first $10k$ optimization steps, keeping the wav2vec 2.0 backbone frozen.
The entire model is then fine-tuned in an end-to-end fashion for the remaining steps.
We use the pre-trained model provided by Baevski et al.\cite{Ot19}\footnote{https://github.com/facebookresearch/fairseq/tree/main/examples/}. Performance for the speaker identification task exceeded 95\% accuracy for all three models whereas the adversarial system delivered a gender recognition accuracy of only 47\%.

\subsection{Gender privacy threat models}
The ability of the systems to conceal the gender information contained in its embeddings is measured by simulating the presence of a third party (an \emph{attacker}) training a 2-layer fully-connected neural network $\mathscr{N}$ to infer the speaker gender from utterance embeddings.  We consider two threat models. In the first one, the attacker is not aware that gender concealment has taken place (\emph{uninformed attack} (uIA)) and therefore trains $\mathscr{N}$ on embeddings that are not gender-protected (in this case, those produced by $M_{s}$ and $M_{sg}$). In the second one, the attacker is aware that model $M_{sga}$ was used to protect the gender identity (\emph{informed attack} (IA)), has access to that model, and trains $\mathscr{N}$ on embeddings produced by that same model. We expect this to result in a more effective attack.

\begin{table}[!t]
\centering
\small
\begin{tabular}{ccc|ccc|}
\cline{4-6}
 &  &  & \multicolumn{3}{c|}{Models} \\ \cline{4-6} 
 &  &  & \multicolumn{1}{c|}{$M_{s}$} & \multicolumn{1}{c|}{$M_{sg}$} & $M_{sga}$ \\ \hline
\multicolumn{1}{|c|}{\multirow{3}{*}{\begin{tabular}[c]{@{}c@{}} EER(\%)\end{tabular}}} & \multicolumn{2}{c|}{Overall} & \multicolumn{1}{c|}{2.36} & \multicolumn{1}{c|}{3.23} & 3.89 \\ \cline{2-6} 
\multicolumn{1}{|c|}{} & \multicolumn{2}{c|}{Male} & \multicolumn{1}{c|}{3.12} & \multicolumn{1}{c|}{4.22} & 4.98 \\ \cline{2-6} 
\multicolumn{1}{|c|}{} & \multicolumn{2}{c|}{Female} & \multicolumn{1}{c|}{3.05} & \multicolumn{1}{c|}{4.21} & 5.26 \\ \hline
\multicolumn{1}{|c|}{\multirow{5}{*}{\begin{tabular}[c]{@{}c@{}} auFDR\end{tabular}}} & \multicolumn{1}{c|}{\multirow{5}{*}{$\alpha$}} & 0 & \multicolumn{1}{c|}{0.98} & \multicolumn{1}{c|}{0.97} & 0.96 \\ \cline{3-6} 
\multicolumn{1}{|c|}{} & \multicolumn{1}{c|}{} & 0.25 & \multicolumn{1}{c|}{0.97} & \multicolumn{1}{c|}{0.97} & 0.95\\ \cline{3-6} 
\multicolumn{1}{|c|}{} & \multicolumn{1}{c|}{} & 0.5 & \multicolumn{1}{c|}{0.97} & \multicolumn{1}{c|}{0.96} & 0.94 \\ \cline{3-6} 
\multicolumn{1}{|c|}{} & \multicolumn{1}{c|}{} & 0.75 & \multicolumn{1}{c|}{0.96} & \multicolumn{1}{c|}{0.95} & 0.92 \\ \cline{3-6} 
\multicolumn{1}{|c|}{} & \multicolumn{1}{c|}{} & 1 & \multicolumn{1}{c|}{0.95} & \multicolumn{1}{c|}{0.94} & 0.91 \\ \hline
\end{tabular}
\vspace{.15cm}
\caption{Performance analysis of the three models for utility and fairness, including EER breakdown by gender and auFDR across various $\alpha$ values (refer to eq.\ref{eq:fdr}) for $\tau$ ranging from 0.1\% to 10\%.}
\label{tab:fair}
\end{table}

\begin{table}[!t]
\centering
\small
\begin{tabular}{|c|c|c|c|}
\cline{2-4}
\multicolumn{1}{c|}{} & \multicolumn{2}{c|}{Data} & \multicolumn{1}{c|}{Attack} \\ 
\cline{2-4}
\multicolumn{1}{c|}{} & Training & Test & AUC (\%) \\ \hline
\multirow{4}{*}{uIA} & $M_s$ & $M_s$ & 97.09 \\ 
\cline{2-4}
 & $M_s$ & $M_{sga}$ & \textbf{46.80} \\ \cline{2-4}
 & $M_{sg}$ & $M_{sg}$ & 98.07 \\ 
\cline{2-4}
 & $M_{sg}$ & $M_{sga}$ & \textbf{40.76} \\ \hline
IA & $M_{sga}$ & $M_{sga}$ & 96.27 \\ \hline
\end{tabular}
\vspace{.15cm}
\caption{Assessment of gender concealment effectiveness under different threat scenarios in terms of AUC.}
\label{tab:privacy}
\end{table}

\begin{figure}[!t]
    \centering
    \includegraphics[width=9cm]{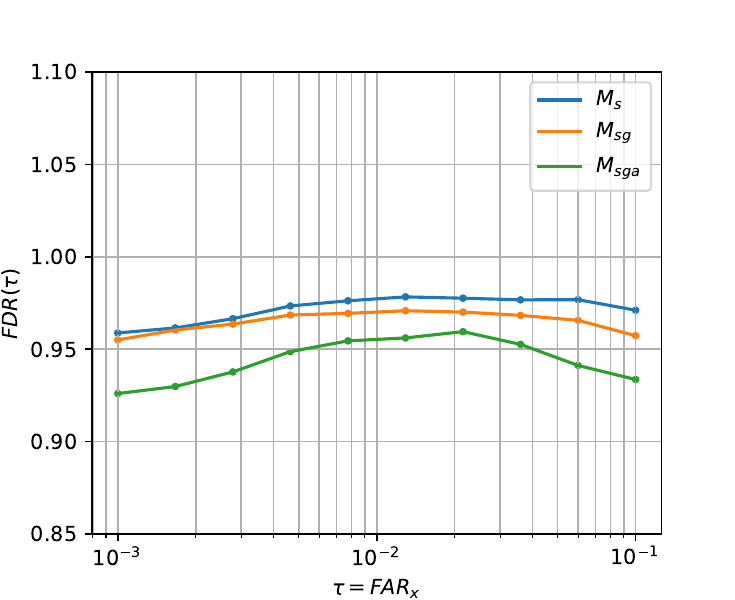}
    \caption{FDR of different ASV systems for different decision thresholds for $\tau$ from 0.1\% to 10\%}
    \label{fig:fdr} 
\end{figure}

\begin{figure*}[!t]
    \centering
    \includegraphics[width=17cm]{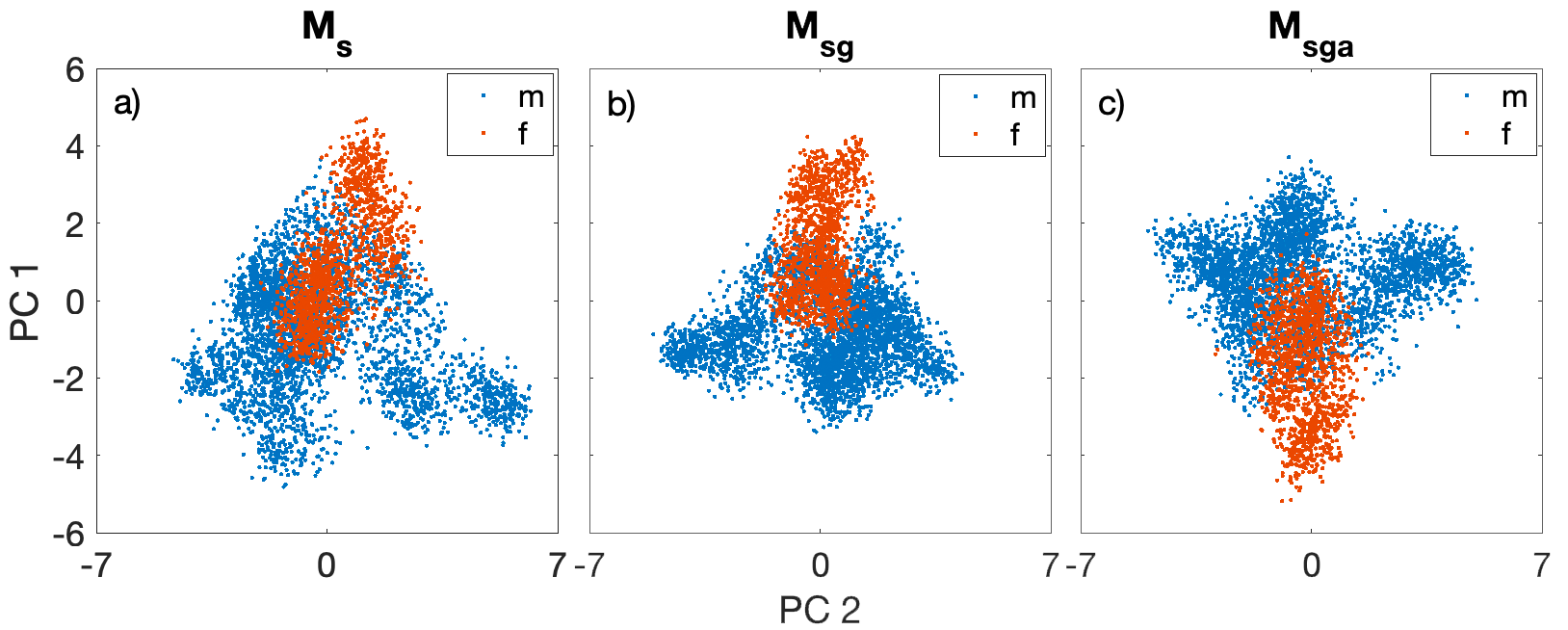}
    \caption{PCA visualizations of features from three models illustrating gender recognition capabilities. Blue points correspond to males and red to females.}
    \label{fig:pca}
\end{figure*}

\begin{figure*}[!t]
    \centering
    \includegraphics[width=17cm]{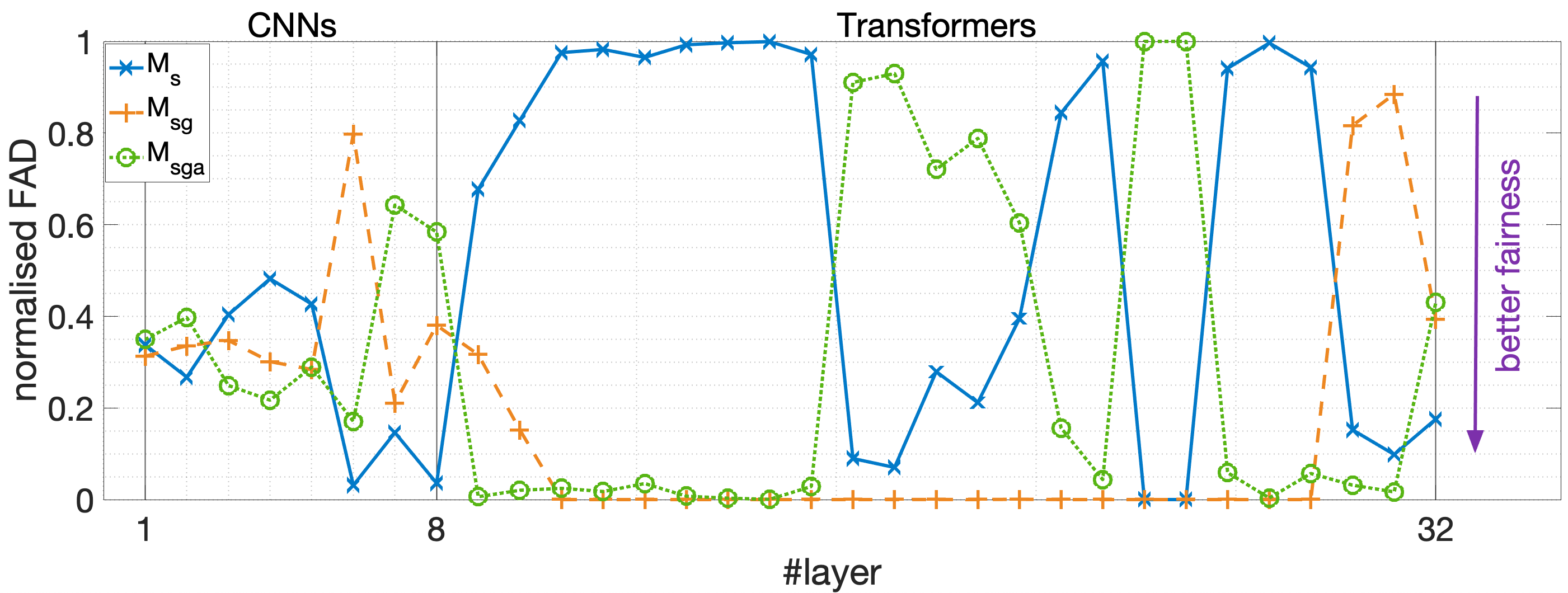}
    \caption{Normalised Fairness Activation Discrepancy (FAD) of different systems at different wav2vec 2.0 module layers. 
    }
    \label{fig:nlnfad} 
\end{figure*}

\section{Experimental results}
We present results for each of the three models $M_s$, $M_{sg}$, and $M_{sga}$. 
Performance is assessed in terms of utility, privacy, and fairness.

In terms of utility, the performance of model $M_s$ is in line with state-of-the-art automatic speaker verification systems, achieving an EER of 2.36\% as shown in Table~\ref{tab:fair}. 
The performance of model $M_{sg}$ and $M_{sga}$ are slightly worse, 3.23\% and 3.89\% respectively, showing that gender influence does not improve speaker recognition. Furthermore, an analysis of the EER broken down by gender shows small differences in speaker recognition for the two genders.

Fairness performances are shown at the bottom of the Table~\ref{tab:fair} in terms of the auFDR for different values of $\alpha$.
All auFDR results are close to 1, indicating reasonable fairness for each group. 
Fig.~\ref{fig:fdr} depicts a plot of the FDR against the threshold for $\alpha = 0.5$. 
Profiles are shown for all three systems. 
The FDR is in all cases above 0.9, and the $M_{s}$ system is always the fairest for each $\tau$. Again, gender influence does not improve fairness.

Privacy performances are presented in Table~\ref{tab:privacy}.
AUC results for uninformed attacks (uIA) are shown at the top. 
When training and testing are performed using embeddings generated using the same, unprotected models, the AUC is 97.09\% and 98.07\% for $M_s$ and $M_{sg}$ models, respectively, demonstrating a lack of privacy protection.
In contrast, when the same uninformed attack is made on the gender-protected model $M_{sga}$, the AUC drops to 46.80\% and 40.76\% respectively.
This significant decrease indicates that the gender classifier predictions become nearly random, successfully concealing the gender information, demonstrating effective protection of privacy.

Performances for the informed attack (IA) are shown in the last row of Table~\ref{tab:privacy}. 
When embeddings are extracted with the $M_{sga}$ model, the AUC is much higher, at 96.27\%. This result underlines the difficulty of obfuscating gender information from embeddings.
Fig.~\ref{fig:pca} reveals an explanation. It illustrates a projection by principal component analysis of the embeddings generated by each of the three models. 
While the $M_{sga}$ model is adversely trained with respect to gender cues, Fig.~\ref{fig:pca}c shows that they persist.
We see that, rather than fully obfuscating gender cues, $M_{sga}$ only rotates the principal components hence why, when trained on similarly-treated training data, gender can still be recognised.

Finally, an analysis of internal bias in terms of FAD has been performed  at different network layers considering male and female groups. This analysis aims to provide insights into the comparative measures of fairness across three distinct models and how they dynamically propagate through the various layers. By examining the internal bias at each layer, we can better understand the impact of model architecture and training data on fairness outcomes. 
As illustrated in Fig.~\ref{fig:nlnfad}, 32 layers were selected in total from the wav2vec 2.0 model. These include 8 layers from the 1D-convolutional encoder and 24 intermediate activation layers from the Transformer modules.

Fig.~\ref{fig:nlnfad} shows the FAD values calculated at different layers. The first layers of the CNNs display similar fairness, likely due to their focus on low-level features.

Contrastingly, Transformer layers, which handle high-level features, have wider fairness variations. $M_s$ and $M_{sga}$ show a complementary behavior as when one achieves high FAD, the other has lower FAD, and vice versa. This could be because $M_s$ was fine-tuned for speaker verification, while $M_{sga}$, with its gradient reversal layer, was trying to suppress gender information. 
As layers progress, all models converge to FAD values, with $M_s$ being the fairest at the end, confirming what is observed in terms of auFDR.

\section{Conclusions and Future Directions}
This research explored the influence of gender information while fine-tuning wav2vec 2.0 for speaker verification. We proposed three models: $M_{s}$, $M_{sg}$, and $M_{sga}$, each with a different focus: speaker recognition, speaker recognition with gender classification, and speaker recognition with gender obfuscation, respectively. Our experiments revealed that $M_{s}$ succeeds in speaker verification (EER of 2.36\%), while $M_{sga}$, designed to hide gender information, performed much worse (EER of 3.89\%). Interestingly, improving gender recognition in the $M_{sg}$ model did not lead to better speaker verification performance (EER of 3.23\%). Privacy evaluations showed effective gender obfuscation against uninformed attacks, but informed attackers could still extract gender information. Fairness evaluations, based on FDR, revealed that highlighting or hiding gender did not significantly impact the fairness of the systems. Furthermore, an analysis of FAD across model layers showed more disparities within Transformer layers, but all systems eventually converged to FAD values that match the auFDR assessment, with system $M_s$ showing superior fairness.

In summary, while we achieved notable results in utility and privacy protection against uninformed attacks, future work includes strengthening gender obfuscation against informed attacks and enhancing fairness across systems.

\section{Acknowledgements}
This work is supported by the TReSPAsS-ETN project funded by the European Union’s Horizon 2020 research and innovation programme under the Marie Skłodowska-Curie grant agreement No. 860813 and partly supported by the VoicePersonae project funded by the French Agence Nationale de la Recherche (ANR) and the Japan Science and Technology Agency (JST).

\vspace{12pt}

\end{document}